\documentclass[a4paper,11pt]{article}
\usepackage{pos}

\newcommand{\ie}{\textit{i.e.}} 
\def\Journal#1#2#3#4{{#1}\,{#2}, #3 (#4);}
\newcommand{\etal}{et al.}
\def\citep#1{\cite{#1}}

\newcommand{\ApJ}{Astrophys. J.}

\newcommand{\PRL}{Phys. Rev. Lett.}
\newcommand{\PRD}{Phys. Rev. D}

\newcommand{\ASR}{Adv. Space Res.}

\renewcommand{\citet}{\cite}
\usepackage{verbatim}

\title{Data driven analysis of Galactic cosmic rays in the heliosphere: diffusion of cosmic protons and nuclei}
\ShortTitle{Cosmic protons and nuclei in the heliosphere}

\author[a]{Nicola Tomassetti}
\author[a]{Bruna Bertucci}
\author[b]{Federico Donnini}
\author[a]{Emanuele Fiandrini}
\author[a]{Maura Graziani}
\author[c]{Behrouz Khiali}
\author[d]{Alejandro Reina Conde}

\affiliation[a]{Department of Physics and Earth's Science, Universit\`a degli Studi di Perugia, Italy}
\affiliation[b]{INFN - Sezione di Perugia, Italy}
\affiliation[c]{INFN - Sezione di Roma Tor Vergata \& ASI Space Science Data Center (SSDC), Roma, Italy}
\affiliation[d]{Instituto de Astrof{\'i}sica de Canarias (IAC), Universidad de La Laguna, Tenerife, Spain}

\abstract{
  Galactic cosmic rays (GCRs) inside the heliosphere are affected by magnetic turbulence and Solar wind disturbances which result in the so-called solar modulation effect. To investigate this phenomenon, we have performed a data-driven analysis of the temporal dependence of the GCR flux over the solar cycle. With a global statistical inference of GCR data collected in space by AMS-02, PAMELA, and CRIS on monthly basis, we have determined the dependence of the GCR diffusion parameters upon time and rigidity. In this conference, we present our results for GCR protons and nuclei, we discuss their interpretation in terms of basic processes of particle transport and their relations with the dynamics of the heliospheric plasma.
} 
\FullConference{36th International Cosmic Ray Conference - ICRC2019 -\\
		July 24th - August 1st, 2019\\
		Madison, WI, U.S.A.}

\begin{document}

\maketitle


\section{Introduction}      
\label{Sec::Introduction}   

When traveling in the heliosphere, energetic charged particles are spatially diffused, magnetically drifted,
advected and decelerated by the solar wind and its embedded magnetic field. 
Due to these effects, the observed energy spectrum of Galactic cosmic rays (GCRs) inside the heliosphere
is significantly different to the local interstellar spectrum (LIS) outside the heliosphere.
Moreover, the modifications of the GCR intensities and energy spectra
are temporal dependent and follows the Sun's variability. 
This phenomenon is referred to as \emph{solar modulation} of GCRs.
Understanding solar modulation is very important in GCR physics,
either to infer the origin of GCRs or to investigate the dynamics of charged particles in the heliospheric turbulence \cite{Moraal2013,Potgieter2013}. 
Modeling the evolution of the GCR radiation in the heliosphere is also important for crewed space missions or for the electronic components
radiation hazard during long-duration missions.
Along with the Voyager-1 data beyond the heliosphere \cite{Cummings2016},
the new precise data from AMS-02 \cite{Aguilar2018PHeVSTime,Aguilar2018LeptonVSTime} and PAMELA \cite{Adriani2013,Martucci2018}
experiments offer a unique possibility to study the solar modulation over a long period of time.

\section{The Numerical Model}  
\label{Sec::Model}             

The propagation of GCRs in the heliosphere is governed by the Parker equation for their phase space density $f(t,R)$\,\cite{Moraal2013}:
\begin{equation}
\label{Eq::Parker}
\frac{\partial f}{\partial t}
=  \nabla\cdot [\mathbf{K}^{S}\cdot\nabla f ]
- (\vec{V}_{sw} + \vec{V}_D) \cdot\nabla f 
+ \frac{1}{3}(\nabla \cdot\vec{V})\frac{\partial f}{\partial (ln R)}  
\end{equation}
where $R=p/Z$ is the rigidity of GCRs (momentum/charge ratio),
$\vec{V}_{sw}$ is the speed of the solar wind, $\vec{V}_{D}$ drift speed,
and $\mathbf{K}^{S}$ is the symmetric component of the GCR diffusion tensor.
The particle flux $J=J(t,R)$ is eventually given by $J=\frac{\beta{c}}{4\pi}n$, where $n=4{\pi}R^{2}f$ is the GCR number density.
In this work, the equation is solved using the \emph{stochastic differential equation} (SDE) method
in steady-state conditions ($\partial/\partial{t}=0$) \cite{Strauss2017},
based on a customized version of the \emph{Solarprop} framework \cite{Kappl2016,Tomassetti2017BCUnc}. 
We implemented a 2D model of heliosphere described by radius $r$ and heliolatitude $\theta$ \citep{Fiandrini2021}.
The heliosphere is modeled as a spherical cavity centered to the Sun from which the wind flows radially.
The wind speed follows a parameterization $V_{sw}(r,\theta,t)$ where, in particular, the latitudinal profile 
is time-dependent, \ie, it evolves with solar activity.
The speed is nearly independent upon helioradius, but it drops to subsonic speeds across the termination shock, at  $r_{\rm{TS}}=85$\,AU,
and then vanishes at the heliopause $r_{\rm{HP}}=122$\,AU. The Earth lies in the equatorial plane, at $r_{0}=$1\,AU from the Sun.
The wind carries a frozen-in Heliospheric Magnetic Field (HMF) which is wounded up in a rotating spiral structure. 
From the solar rotation with a characteristic tilt angle $\alpha$ between magnetic and rotational axis,
a waving Heliospheric Current Sheet (HCS) is generated. 
The HCS is a rotating structure which divides the HMF into two hemispheres of opposite polarity.
The $\alpha$-angle is measured real-time by the  Wilcox Solar Observatory (WSO), since the 70's, on a 10-day basis \cite{Hoeksema1995}.
It ranges from $\sim{0-10}^{\circ}$ during solar minimum (flat HCS) to $\sim{80-90}^{\circ}$ during maximum and reversal (wavy HCS).

In the propagation of GCRs in the HMF, various processes occurring at different spatial scales:
diffusion arises from the erratic random-walk scattering of the particles off the small-scale HMF turbulence.
Drift is due to the large-scale regular component of the HMF, from spatial gradient, curvature, and in proximity of the HCS.
Diffusion and drift are included in the symmetric and antisymmetric parts of the diffusion tensor $\mathbf{K}$, respectively:
$\mathbf{K}=\mathbf{K}^S+\mathbf{K}^A$, with $K_{ij}^S = K_{ji}^S$ and $K_{ij}^A = -K_{ji}^A$.
The $\mathbf{K}^{S}$ tensor can be also divided parallel and perpendicular diffusion $K_{\parallel}$ and $K_{\perp}$,
or terms of the mean free paths $\lambda_{\parallel}$ and $\lambda_{\perp}$, such that
$K_{\parallel} = \beta c \lambda_{\parallel}/3$,  where $\beta=v/c$ is the particle speed.
The perpendicular diffusion length $\lambda_{\perp}$
is assumed proportional to the parallel one, $\lambda_{\perp}= \xi \lambda_{\parallel}$, with  $\xi \cong\,0.02$ \cite{Giacalone1999}.
The rigidity dependence of the GCR diffusion coefficients arises from the
cyclotron resonance condition of GCR scattering on the HMF irregularities, occurring when the Larmor radius $r_{L}=r_{L}(R)$ 
is comparable with the spatial scale size of the irregularities $\hat{\lambda}$.
From the condition $r_{L} \sim \hat{\lambda}$, it turns out that GCRs with rigidity $R$ resonate at wave number $k_{\rm{res}} \sim 1/R$.
The spatial scale of irregularities however follows a turbulence spectrum of the type $w(k) \propto k^{-\eta}$, 
in terms of wave number $k=2\pi/\lambda$.
The index $\eta$ depends on type and spatial scales of the turbulence energy cascade.
Thus, $\lambda_{\parallel}$ will depend on rigidity as $\lambda_{\parallel} \sim R^{2-\eta}$.
On a wide range of scale, various regimes can be distinguished for the HMF power spectrum \cite{Kiyani2015}. 
A good parameterization for the rigidity dependence of $\lambda_{\parallel}$
is the \emph{double power-law} function, defined by two spectral indices $a$ and $b$ and a critical rigidity value $R_{k}$ \cite{Potgieter2013}.
For $K_{\parallel}$ we have adopted the following description: 
\begin{equation}\label{Eq::Par_diff}
K_{\parallel} = \frac{K_{0}}{3}\beta \left(\frac{B_0}{B}\right)   \left(\frac{R_0}{R}\right)^a 
 \times \left[ \frac{(R/R_0)^h + (R_k/R_0)^h }{1 + (R_k/R_0)^h} \right]^{\frac{b-a}{h}} 
\end{equation}
where $K_{0}$ is a constant in units of $10^{23}$ $cm^2 s^{-1}$ and $R_{0}\equiv$\,1\,GV sets the rigidity scale.
The the HMF magnitude is $B$ while $B_{0}$ is the \emph{local} field value at $r_{0}=$\,1 AU. 
The parameters $a$ and $b$ set the two slopes of the rigidity dependence below and above $R_{k}$, respectively.
The smoothness of the transition is regulated by the parameter $h$.
The perpendicular mean free path follows from $\lambda_{\perp}= \xi \lambda_{\parallel}$, with the 
addition of polar corrections \cite{Heber1998}.

The parameters regulating GCR diffusion are subjected to temporal evolution following the Solar Cycle \cite{Manuel2014}. T
The diffusion parameter set ${K_{0}, a, b}$ their temporal evolution is determined by a global fit to the monthly data of
AMS-02 and PAMELA  \cite{Aguilar2018PHeVSTime,Adriani2013,Martucci2018}.
%
\begin{figure*}[ht!]
\centering
\includegraphics[width=0.85\textwidth,scale=0.45]{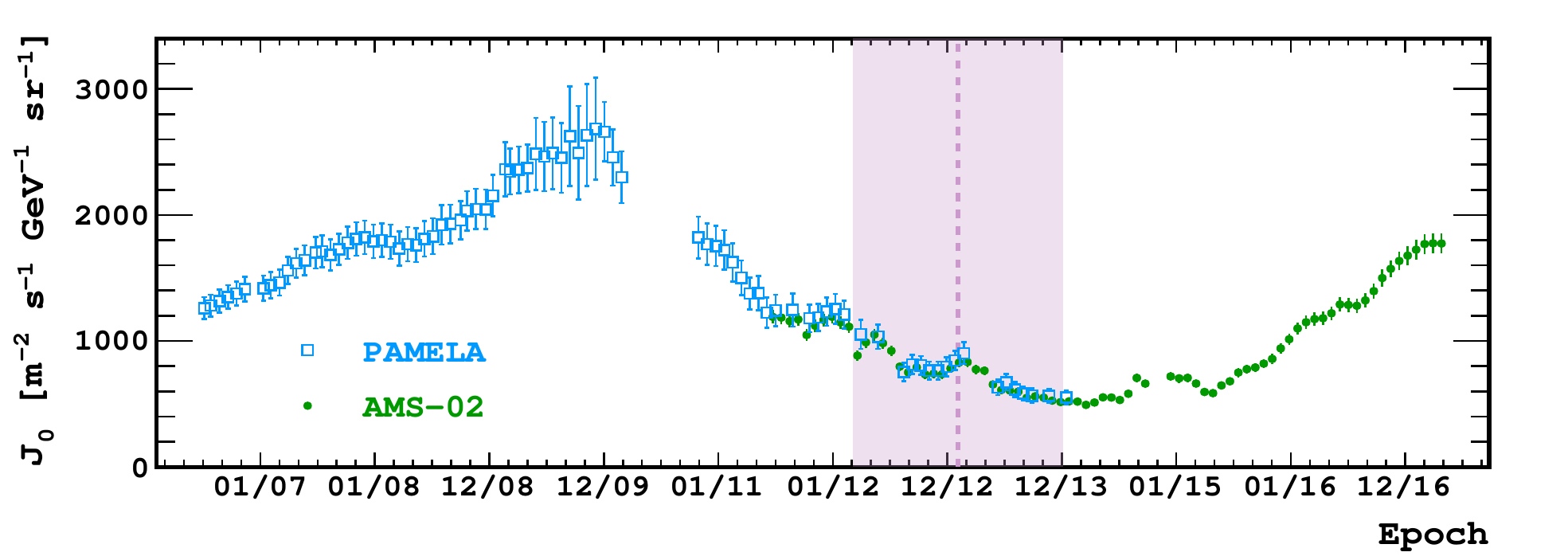}
\caption{\footnotesize{BR averaged flux $J_{0}$ evaluated in the reference energy range between 0.49-0.62 GeV
  from PAMELA (open squares)\citep{Adriani2013Protons,Martucci2018}
  and AMS-02 (filled circles)  \citep{Aguilar2018PHeVSTime,Aguilar2018LeptonVSTime}.
  The vertical dashed line shows the epoch of the HMF polarity inversion, along with the shaded area indicating the reversal epoch.}}
\label{Fig::ReferenceFlux}       
\end{figure*}
%
%
The intensity of the GCR proton fluxes in the energy range between 0.49 - 0.62 GeV
are shown in Fig.\,\ref{Fig::ReferenceFlux} as a function of time for both the PAMELA and AMS-02 data sets. 
Along with the three GCR \emph{diffusion parameters}, we identify a set of three \emph{heliospheric parameters} that describe the
status of the modulation region at a given epoch. They are the HCS tilt angle $\alpha$, the local value of the HMF $B_{0}$,
and the magnetic polarity $A$, where the latter is defined as the sign of the Sun's magnetic field in the outgoing direction from its North pole.
The parameter set ${\alpha, B_{0}, A}$ is also time dependendent.
Finally, to compute the modulation according to  Eq.\,\ref{Eq::Parker}, an input LIS model should be specified as boundary condition.
Models of LIS include Galactic astrophysics processes such as acceleration and interstellar propagation.
To compute the GCR proton LIS, we employ calculations from recent works \cite{Tomassetti2015TwoHalo,Tomassetti2012Hardening,Feng2016,Tomassetti2018PHeVSTime}.
Our proton LIS was tight constrained with low-energy interstellar data from Voyager-1 at $\sim$\,100\,-500\,MeV of kinetic energy \cite{Cummings2016},
and with AMS-02 high-energy data at $E\gtrsim$\,100\,GeV \cite{Aguilar2018PHeVSTime,Aguilar2015Proton,Aguilar2015Helium}.
The resulting proton LIS agrees fairly well with other recent models
\cite{Boschini2017,Corti2019,Tomassetti2017TimeLag,Tomassetti2015PHeAnomaly,Tomassetti2017Universality}.

\section{The parameter extraction}   
\label{Sec::Model}                   

We model the time-dependence of the problem 
by making use of a continuous series of equilibrium solution of Eq.\,\ref{Eq::Parker},
where each solution is obtained for a given set of six input parameters.
The three \emph{heliospheric parameters} are obtained using a backward moving average (BMA) of
observations by WSO observatory and by \emph{in situ} measurements of the ACE space probe.
For a given epochs $t$, the average is calculated within a time window $[t-\Delta{T}, t]$, with  $\Delta{T}=6-12$\,months. 
The window is chosen so that the BMA values of $\hat{\alpha}$ (from WSO) and $\hat{B}_{0}$ (from ACE) reflect
the average HMF conditions sampled by GCRs arriving Earth \cite{Fiandrini2021,Tomassetti2017TimeLag}. 
The remaining \emph{diffusion parameters} $K_{0}$, $a$, and $b$ have been determined with a global fit on the GCR proton measurements from AMS-02 and PAMELA.
To fit the GCR data, we have built a six-dimensional grid. Each nodes of the grid corresponds to a configuration of the
vector $\vec{q}=$ ($\alpha$, $B_0$, $A$, $K_0$,  $a$, $b$). The grid has a total number of 938,400 nodes.
Using the stochastic technique, the GCR proton spectrum $J_{m}(E, \vec{q})$ was evaluated for each node of the grid
at kinetic energies from 20 MeV to 200 GeV.
This task required the simulation of 14 billion trajectories, corresponding to several months of CPU time. 
For each trajectory, the pseudoparticles were backwardly-propagated from Earth to the heliopause and then re-weighted according to the LIS.
Once the proton grid has completed, the parameters were inferred using the GCR proton data.
Using the measured fluxes $J_{d}(E,t)$ made at epoch $t$, the model calculation $J(E,\vec{q})$ with the
heliospheric parameters ${\hat{\alpha},\hat{B_{0}},\hat{A}}$ fixed by the BMA procedure,
a global $\chi^{2}$ function was calculated as follows:
\begin{equation} \label{Eq::ChiSquare}
  \chi^{2}(K_{0},a,b)= \sum_{i}  \frac{\left[ J_{d}(E_{i},t) - J_{m}(E_{i}, \vec{q}) \right]^{2}}{\sigma^{2}(E_{i},t)}
\end{equation} 
The best-fit diffusion parameters were then obtained by the minimization of the $\chi^{2}$ function.
In Eq.\,\ref{Eq::ChiSquare}, the errors are given by $\sigma^{2}(E_{i},t) = \sigma_{d}^{2}(E_{i},t) + \sigma_{mod}^{2}(E_{i},t)$, \ie,
by the sum in quadrature of several contributions: experimental uncertainties in the data, theoretical uncertainties
of the model, and errors associated with the minimization procedure.
\begin{figure*}[hbt]
\centering
\includegraphics[width=0.85\textwidth,scale=0.50]{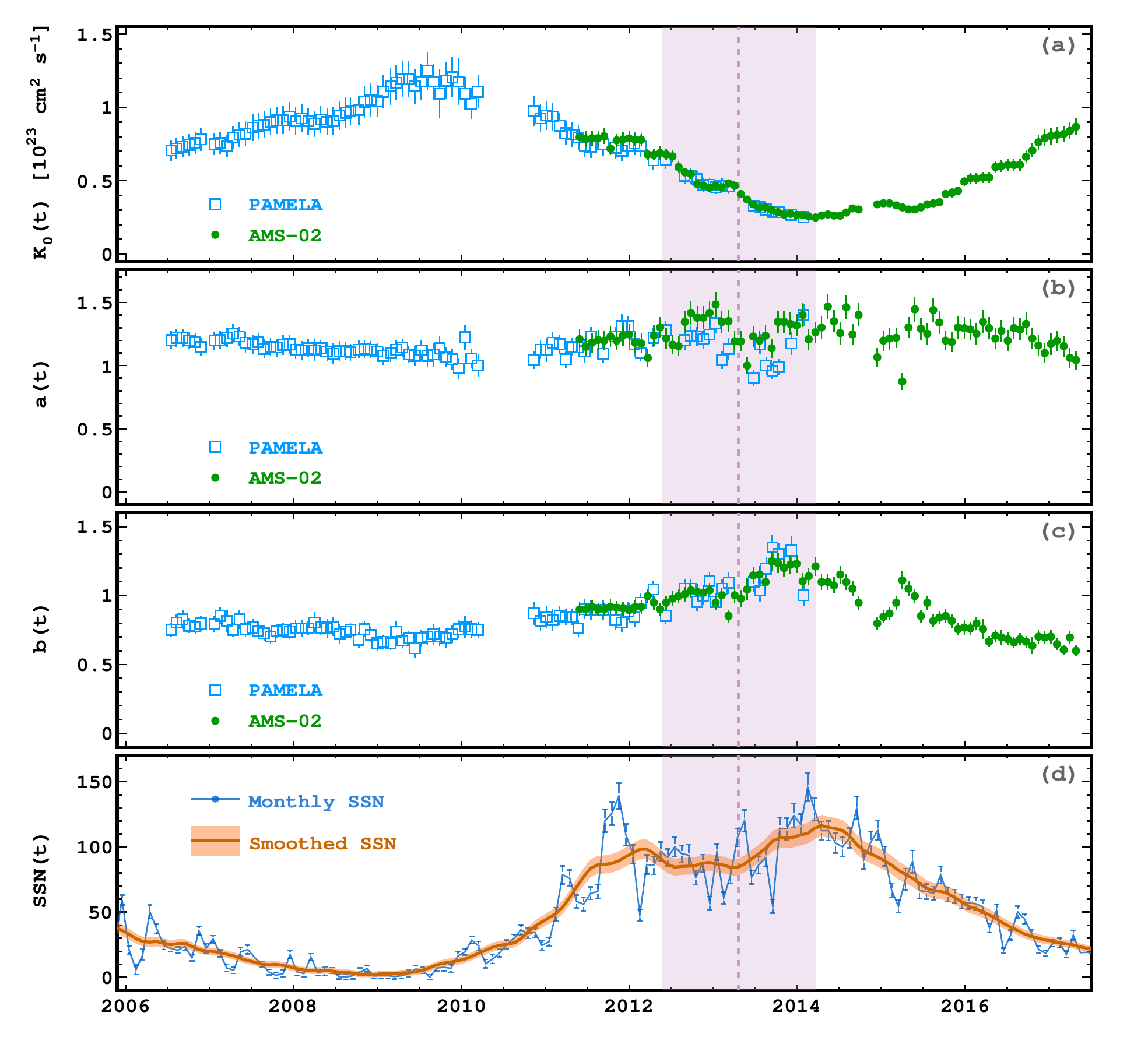} 
\caption{\footnotesize{Results for the best-fit model parameters $K_{0}$, $a$, and $b$ determined using the time-resolved proton flux measurements
  from PAMELA (open squared) and AMS-02 (filled circles). 
  In panel (d), the monthly averaged and smoothed SSN is shown. The vertical dashed line indicates the reversal epoch $T_{\rm{rev}}$
  and the shaded area around it shows the transition epoch where the HMF polarity is weakly defined.}}
\label{Fig::BestFitParametersVSTime}
\end{figure*}

\section{Results and discussion}  
\label{Sec::Results}              
%
%
%
\begin{figure*}[ht!]
\centering
\includegraphics[width=0.85\textwidth,scale=0.45]{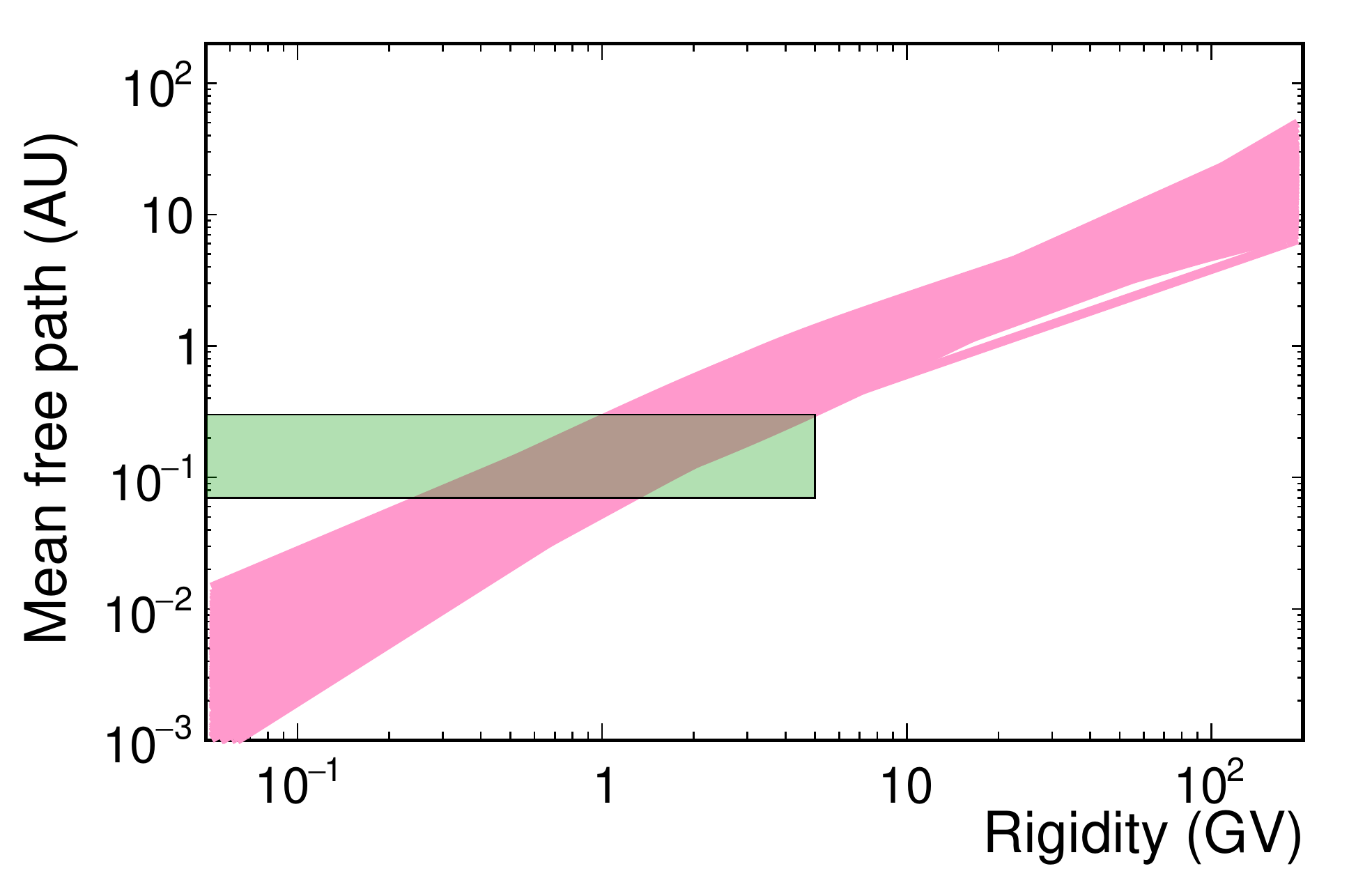}
\caption{\footnotesize{Envelope of the diffusion mean free paths $\lambda_{\parallel}$ as function of GCR rigidity inferred in the examined period (pink band).
  The shaded green box corresponds to the Palmer consensus for reference to observational data on $\lambda_{\parallel}$ \citep{Palmer1982}.}}
\label{Fig::MeanFreePath}       
\end{figure*}
For the diffusion parameters, our  least-square minimization procedure returned a time-series of
best-fit values and their corresponding uncertainties. The results of the fitting are shown in Fig.\,\ref{Fig::BestFitParametersVSTime}.
In the figure, we plot the temporal dependence of the parameters $K_{0}(t)$ (a), $a(t)$ (b), $b(t)$ (c), along with
the corresponding evolution of the monthly/smoothed sunspot number (SSN) (d) as a proxy of the solar activity cycle. 
The color codes represent the data used to make the fits, \ie, the time-series of GCR fluxes from AMS-02 (green filled circles) and PAMELA (blue open suqares).
The considered period covered a significant fraction the Solar Cycle, including the magnetic reversal phase around $T=T_{\rm{rev}}$, indicated by the shaded band,
where the HMF polarity $A$ switched from positive to negative.
From the figure, it can be seen that the diffusion parameters show a remarkable temporal dependence, and such a dependence is well correlated with solar activity.
The diffusion normalization parameter $K_{0}$  shows a clear temporal dependence and a marked anti-correlation with the monthly SSN.
The parameter appears to be maximum in the $A<0$ epoch before
reversal ($t\ll{T_{\rm{rev}}}$), and in particular during the long solar minimum of 2009-2010.
The minimum of $K_{0}$ is reached during solar maximum of 2014, about one year after the reversal.
Physically, larger $K_{0}$ values imply faster GCR diffusion, thereby causing a minor modification of the LIS, \ie, a higher GCR flux at the GeV scale.
In contrast, lower $K_{0}$ values imply slower diffusion which and a stronger attenuation of the GeV flux.
This behavior can be interpreted within the Force-Field model where, in fact, one has $\phi\propto 1/K_{0}$ \citep{Tomassetti2017BCUnc}.
Within the  Force-Field model, the parameter $\phi$ is interpreted as the average kinetic energy loss of GCR protons inside the heliosphere.
Thus, one expects a positive correlation between $K_{0}(t)$ of Fig.\,\ref{Fig::BestFitParametersVSTime} and the
reference GCR flux $J_{0}=J(t,E)$ of Fig.\,\ref{Fig::ReferenceFlux}.
Our finding are in agreement with earlier works \citep{Manuel2014,Tomassetti2017TimeLag,Corti2019}. 
Interestingly, the diffusion index $b$ shows a distinct time dependence, while the index $a$ has milder variations.
This suggests that the turbulence spectrum in the inertial range
evolve as a function of the solar activity, with a clear delayed peak at the solar maximum.
The inferred spectral index of the turbulence in the energy-containing range is about $\nu_{ec}=0.79\pm0.13$ in the examined period.
In the inertial range, the index evolves from $\nu_{in}= 0.74\pm0.08$ at solar minimum to $\approx{1.3}\pm0.15$ during the solar maximum.
In both ranges, our parameters are in agreement with the measured slopes of the HMF power spectrum on Jan-Feb 2007 \cite{Kiyani2015}.
In most of numerical models of solar modulation, these parameters are usually assumed to be time-independent.
Variations in these parameters imply changes in the HMF turbulence spectrum \cite{Usoskin2019,Horbury2005}. 
Figure\,\ref{Fig::MeanFreePath} shows the envelope of the mean free paths $\lambda_{\parallel}$ for parallel diffusion inferred in the examined period.
It can be seen that our results are in agreement with the Palmer consensus, \ie, the
large collection of observational measurements on the scattering mean free path \citep{Palmer1982}.

\section{Acknowledgements}  
\label{Sec::Results}        

We acknowledge the support of ASI
under agreement \emph{ASI-UniPG 2019-2-HH.0}.


\end{document}